\documentclass[journal]{IEEEtran}
\usepackage[T1]{fontenc}
\usepackage[utf8]{inputenc}  
\usepackage{cite}
\usepackage{amsmath,amssymb,amsfonts}
\usepackage{algorithmic}
\usepackage{graphicx}
\usepackage{textcomp} 
\usepackage{subfigure}
\usepackage{newtxtext} 
\usepackage{newtxmath} 
\usepackage{bm}
\usepackage{hyperref}
\usepackage{url}
\usepackage{float}
\usepackage{multirow}
\usepackage{booktabs}
\newcommand{\D}{\mathrm{d}}
\newcommand{\e}{\mathrm{e}}
\newcommand{\T}{\top}

\newcommand{\J}{\mathbf{J}}

\newcommand{\Rs}{R_\mathrm{s}}
\newcommand{\Ld}{L_\mathrm{d}}
\newcommand{\Lq}{L_\mathrm{q}}

\newcommand{\Gs}{\bm{\Gamma}_\mathrm{\! s}}

\newcommand{\psisal}{\psi_\alphaup}
\newcommand{\psisbe}{\psi_\betaup}
\newcommand{\isal}{i_\alphaup}
\newcommand{\isbe}{i_\betaup}
\newcommand{\usal}{u_\alphaup}
\newcommand{\usbe}{u_\betaup}

\newcommand{\psisd}{\psi_\mathrm{d}}
\newcommand{\psisq}{\psi_\mathrm{q}}
\newcommand{\isd}{i_\mathrm{d}}
\newcommand{\isq}{i_\mathrm{q}}
\newcommand{\usd}{u_\mathrm{d}}
\newcommand{\usq}{u_\mathrm{q}}

\newcommand{\us}{\mathbf{u}_\mathrm{s}}
\newcommand{\is}{\mathbf{i}_\mathrm{s}}
\newcommand{\psis}{\bm{\psiup}_\mathrm{s}}
\newcommand{\uss}{\mathbf{u}_\mathrm{s}^\mathrm{s}}
\newcommand{\iss}{\mathbf{i}_\mathrm{s}^\mathrm{s}}
\newcommand{\psiss}{\bm{\psiup}_\mathrm{s}^\mathrm{s}}
\newcommand{\psif}{\bm{\psiup}_\mathrm{f}}

\newcommand{\isk}{\mathbf{i}_{\mathrm{s}\ell}}
\newcommand{\hatisk}{\hat{\mathbf{i}}_{\mathrm{s}\ell}}
\newcommand{\psisk}{\bm{\psiup}_{\mathrm{s}\ell}}
\newcommand{\hatpsisk}{\hat{\bm{\psiup}}_{\mathrm{s}\ell}}

\newcommand{\omegam}{\omega_\mathrm{m}}
\newcommand{\thetam}{\vartheta_\mathrm{m}}
\newcommand{\taum}{\tau_\mathrm{m}}

\newcommand{\Wsym}{\overline{W}}
\newcommand{\Ws}{W^\mathrm{s}}
\newcommand{\Wper}{\tilde{W}}
\newcommand{\xper}{\tilde{\mathbf{x}}}

\newcommand{\thetaper}{\bm{\varthetaup}}
\newcommand{\tauper}{\bm{\tauup}}
\newcommand{\g}{\mathbf{g}}
\newcommand{\Conj}{\mathbf{C}}
\graphicspath{{figs/}}  
 
\usepackage[dvipsnames]{xcolor}

\begin{document}

\title{Gradient Networks for Magnetic Modeling of Synchronous Machines}

\author{Junyi Li, \IEEEmembership{Graduate Student Member,~IEEE}, Tim Foi\ss ner, Floran Martin,\\ Antti Piippo, and Marko Hinkkanen, \IEEEmembership{Fellow, IEEE}\vspace{-.2em}
\thanks{This work has been submitted to the IEEE for possible publication. Copyright may be transferred without notice, after which this version may no longer be accessible.}
\thanks{This work was supported in part by ABB Oy, in part by the Aalto University House of AI, and in part by the Research Council of Finland Centre of Excellence in High-Speed Electromechanical Energy Conversion Systems. The authors acknowledge the use of EPE infrastructure of Aalto School of Electrical Engineering.}
\thanks{Junyi Li, Floran Martin, and Marko Hinkkanen~are with the Department of Electrical Engineering and Automation,~Aalto University, 02150 Espoo, Finland (e-mail: junyi.li@aalto.fi; floran.martin@aalto.fi; marko.hinkkanen@aalto.fi).}
\thanks{Tim Foi\ss ner and Antti Piippo are with ABB Oy, Drives, 00380 Helsinki, Finland (e-mail: tim.foissner@fi.abb.com; antti.piippo@fi.abb.com).}}

\maketitle

\begin{abstract}
This paper proposes a magnetic model for saturable synchronous machines, including spatial harmonics. Black-box neural networks do not guarantee reciprocity or invertibility of the modeled magnetic characteristics. In the proposed model, a monotone gradient network represents the gradient of the magnetic field energy, from which the stator current and the electromagnetic torque are obtained. Hence, the model inherently satisfies reciprocity and energy conservation, its outputs are smooth, and the map between the flux linkages and currents is invertible. A dual model based on the co-energy is also given. An activation function that is computationally more efficient than the softmax activation is proposed for real-time control. The model is validated using measured data from two permanent-magnet (PM) synchronous reluctance machines (5.6~kW and 28~kW) and finite-element method (FEM) data. For the 5.6-kW machine, a model with 41 parameters reproduces the measured current and flux-linkage maps with rms errors of about 0.02~p.u.\ and 0.01~p.u., respectively, when 24 measured points are used for training. The model is also demonstrated in real-time closed-loop control at a sampling frequency of 8~kHz.
\end{abstract}


\section{Introduction}
\IEEEPARstart{D}{ynamic} models of synchronous machines are essential for control, estimation, monitoring, and optimization. A particularly challenging aspect is the magnetic model, which describes the relationship between flux linkages, currents, rotor angle, and electromagnetic torque \cite{Woo1968, Fit2003}. In permanent-magnet (PM) and synchronous reluctance machines, magnetic saturation and cross-saturation are significant and must often be included in control models, which run on embedded processors with limited memory and computation time. Spatial harmonics, on the other hand, can be included in models for time-domain simulations during the design stage.

The nonlinear magnetic behavior is typically represented by analytical functions~\cite{Qu2012, Gar2012, Su2022, Var2023, Lel2024}, semi-analytical models~\cite{Vag1999}, lookup tables~\cite{Che2015, Li2017, Lee2021, Fer2021, Boj2026}, or neural networks~\cite{Jas2021, Pas2023, Liu2024, Lee2025, Sha2026}. These models can be characterized from finite-element method (FEM) data~\cite{Siz2012}, laboratory measurements~\cite{Arm2013,Boj2026}, or automatic identification routines~\cite{Hin2017}. Analytical models are limited by their functional form and are difficult to extend to higher dimensions (e.g., for spatial harmonics or multi-phase machines). Lookup tables require dense grids and large memory, especially in higher dimensions, and linear interpolation yields discontinuous derivatives. In $\mathrm{dq}\theta$ lookup tables~\cite{Fer2021, Boj2026}, the flux linkages are also tabulated as functions of the rotor angle. The torque can then be computed from the co-energy, which is obtained by integrating the tabulated flux linkages~\cite{Boj2026}, but the tables themselves are not constrained to be reciprocal~\cite{Sau1992} or monotone. Black-box neural networks~\cite{Pas2023, Liu2024, Lee2025} can approximate high-dimensional complex maps and require less memory than lookup tables, but they may require large training datasets and do not guarantee reciprocity or invertibility. Although some architectures incorporate spatial periodicity~\cite{Jas2021, Sha2026}, neither reciprocity nor monotonicity is guaranteed.

Physics-informed neural networks address these limitations by embedding known physical principles~\cite{Rai2019, Cra2020, Gre2019, Des2021}. Electric machines are port-Hamiltonian systems~\cite{Sch2014}, in which the current and electromagnetic torque are gradients of the magnetic field energy with respect to the flux linkage and rotor angle, respectively~\cite{Woo1968, Fit2003}. Hamiltonian neural networks~\cite{Gre2019, Des2021} could exploit this structure by modeling the energy as a scalar network and differentiating it, which improves data efficiency and physical consistency. However, a close approximation of the energy does not guarantee a close approximation of its gradient, and these networks become harder to optimize as the input dimension grows~\cite{Cha2025}. Furthermore, the network has to be differentiated during both training and evaluation. The convexity of the energy function could be guaranteed using input-convex neural networks \cite{Amo2017}, but the differentiation is still needed. A similar differentiation principle was applied to electromagnetic torque modeling~\cite{Sha2026}, but the relationship between currents and flux linkages was not considered.

In this paper, the magnetic characteristics of synchronous machines~\cite{Woo1968, Fit2003} are modeled using monotone gradient networks~\cite{Cha2025}. Gradient networks directly model the gradient of a convex function, without differentiating a scalar network. The main contributions are:
\begin{enumerate}
    \item The stator current and electromagnetic torque are modeled directly as the gradient of the field energy using a monotone gradient network~\cite{Cha2025} (Section~\ref{sec:architectures}). Reciprocity, strong monotonicity, and thus invertibility hold for any input and for any values of the learnable parameters. A dual formulation based on the co-energy gives the flux linkage and torque with the same properties.
    \item Spatial harmonics are included by means of a periodic representation of the rotor angle, so that the same network gives the current (or the flux linkage) and the torque with its ripple. It is shown that the monotonicity of the network with respect to this representation does not restrict the model. The q-axis symmetry is enforced by symmetrizing the energy function (Sections~\ref{sec:sm_model_rot} and~\ref{sec:architectures}).
    \item A $p$-norm gradient activation without transcendental functions is proposed as a computationally efficient alternative to the softmax activation, whose universal approximation property is proved in~\cite{Cha2025} (Section~\ref{sec:gradnet}). Guidelines are given for selecting the activation functions for the current and flux-linkage maps, taking into account the behavior outside the data range.
    \item The models are validated using measured data of two PM synchronous reluctance machines (5.6~kW and 28~kW) and FEM data of the 5.6-kW machine (Section~\ref{sec:results}). A model trained using 24 measured points is compared with a lookup table, an analytical model, and other neural networks. A model with 23 parameters is used in the control law, flux observer, and reference generation of a real-time control system.
\end{enumerate}

This paper is organized as follows. Section~\ref{sec:sm_model} presents the machine model, including the periodic representation of the rotor angle and the q-axis symmetry. Section~\ref{sec:proposed_model} details the proposed architecture. Section~\ref{sec:results} presents offline validation and real-time control results. Section~\ref{sec:conclusions} concludes the paper.

\begin{figure}[t]
  \centering
  \setlength{\subfigcapskip}{4pt}
  \subfigure[]{
    \includegraphics{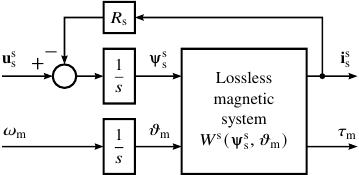}}
  \hfill
  \subfigure[]{
    \includegraphics{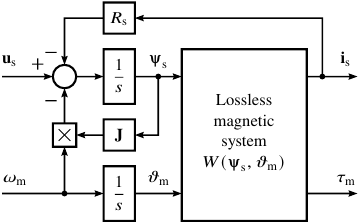}}
  \caption{Electromechanical dynamics of a generic synchronous machine: (a) stator coordinates; (b) rotor coordinates. The blocks $1/s$ denote integration in time.}
  \label{fig:sm_block_both}
\end{figure}

\section{Physics-Based Machine Model} \label{sec:sm_model}
This section describes the fundamental dynamic model of a generic synchronous machine \cite{Woo1968, Fit2003}. Scalars are denoted by italic letters, column vectors by bold lowercase letters, and matrices by bold uppercase letters. All quantities are in per-unit unless otherwise specified.

\subsection{Stator Coordinates}
The synchronous machine model in stator ($\alphaup\betaup$) coordinates is first considered. The state equations are  
\begin{subequations} \label{eq:dyn_stat}
\begin{align} \label{eq:dpsiss}
  \frac{\D \psiss}{\D t} &= \uss - \Rs \iss \\
  \frac{\D \thetam}{\D t} &= \omegam 
\end{align} 
\end{subequations}
where $\psiss = [\psisal,\psisbe]^\T$ is the stator flux-linkage vector, $\iss = [\isal, \isbe]^\T$ is the stator current vector, $\uss = [\usal, \usbe]^\T$ is the stator voltage vector, and $\Rs$ is the stator resistance. Furthermore, $\thetam$ is the electrical angle of the rotor d-axis with respect to the stator coordinates, and $\omegam$ is the electrical angular speed of the rotor.

Assuming a lossless (conservative) magnetic field system, the magnetic behavior is fully described by its field energy function $\Ws(\psiss, \thetam)$. Hence, the stator current and electromagnetic torque are given by \cite{Woo1968, Fit2003} 
\begin{subequations} \label{eq:mag_stat}
\begin{align}
  \iss &= \left[\frac{\partial \Ws(\psiss,\thetam)}{\partial \psiss} \right]^\T \\
  \taum &= -\frac{\partial \Ws(\psiss,\thetam)}{\partial \thetam} .
\end{align}
\end{subequations} 
The minus sign appears in the torque expression since the positive mechanical power is defined out of the machine. It can be seen that the current and electromagnetic torque (with negative sign) constitute the gradient of the field energy function \cite{Sch2014, Boy2004}. Fig.~\ref{fig:sm_block_both}(a) shows the block diagram corresponding to \eqref{eq:dyn_stat} and \eqref{eq:mag_stat}.

If needed, core losses can be included outside the magnetic model by adding an equivalent core-loss resistance (or a more general nonlinear dissipative element) in parallel to the main magnetic circuit, as is standard practice \cite{Vag1999, Che2015}. This reduces the current entering the lossless part of the model, but otherwise the model remains unchanged. Neither the core losses nor magnetic hysteresis are included in the models of this paper.

\subsection{Rotor Coordinates} \label{sec:sm_model_rot}
The machine model is more convenient to express in rotor (dq) coordinates, which rotate with the rotor d-axis. Coordinate transformation to rotor coordinates can be expressed using the matrix exponential as
\begin{align} \label{eq:coord_transform}
  \psis = 
  \begin{bmatrix}
    \cos\thetam & \sin\thetam \\
    -\sin\thetam & \cos\thetam
  \end{bmatrix} \psiss = \e^{-\thetam\J}\psiss
\end{align}
where $\J = [\begin{smallmatrix} 0 & -1 \\ 1 & 0 \end{smallmatrix}]$ is the orthogonal rotation matrix. The flux-linkage vector $\psis = [\psisd,\psisq]^\T$ is used as an example, but other vectors are transformed similarly.  

\subsubsection{Model Structure} 
Using \eqref{eq:coord_transform}, the state equations \eqref{eq:dyn_stat} are transformed to rotor coordinates as
\begin{subequations} \label{eq:dyn}
\begin{align} \label{eq:dpsis}
  \frac{\D \psis}{\D t} &= \us - \Rs \is - \omegam \J \psis \\ \label{eq:dthetam}
  \frac{\D \thetam}{\D t} &= \omegam 
\end{align}
\end{subequations}
where $\is = [\isd, \isq]^\T$ is the current vector and $\us = [\usd, \usq]^\T$ is the voltage vector. The field energy function can be expressed in rotor coordinates, $W(\psis, \thetam) = \Ws(\psiss, \thetam)$. Applying the coordinate transformation and the chain rule (see Appendix~\ref{app:transformation}), the magnetic model \eqref{eq:mag_stat} becomes
\begin{subequations} \label{eq:mag}
\begin{align} \label{eq:mag_a}
  \is &= \left[\frac{\partial W(\psis,\thetam)}{\partial \psis}\right]^\T \\
  \taum &= \is^\T\J\psis - \frac{\partial W(\psis,\thetam)}{\partial \thetam} .
\end{align}
\end{subequations}
The gradient relation \eqref{eq:mag_a} defines the current map $\is(\psis, \thetam)$. Fig.~\ref{fig:sm_block_both}(b) shows the corresponding block diagram.

In magnetic systems, increasing flux linkage requires increasing current \cite{Woo1968}. Hence, the incremental inverse inductance matrix $\Gs = \partial \is / \partial \psis$ is positive definite, which makes the current map strictly monotone. The field energy $W(\psis, \thetam)$ is thus strictly convex with respect to $\psis$, and the current map is one-to-one. 

If spatial harmonics are omitted, the field energy depends only on the flux linkage, $W(\psis)$, simplifying the current map to $\is(\psis)$ and the electromagnetic torque to $\taum = \is^\T\J\psis$. As a familiar special case, this formulation includes the standard magnetically linear machine, where the field energy is quadratic, $W(\psis) = \frac{1}{2}(\psis - \psif)^\T \Gs (\psis - \psif)$, with the constant inverse inductance matrix $\Gs = \mathrm{diag}(1/\Ld, 1/\Lq)$ and the PM-flux vector $\psif = [\psi_\mathrm{f}, 0]^\T$. In this case, \eqref{eq:mag_a} reduces to the linear relationship $\is = \Gs (\psis - \psif)$.

\subsubsection{Co-Energy-Based Model}
Alternatively, a dual model can be derived via the Legendre transform $W(\psis, \thetam) + W'(\is, \thetam) = \is^\T\psis$, which relates the field energy $W$ and co-energy $W'$ \cite{Woo1968,Fit2003}. This yields
\begin{subequations} \label{eq:dual_mag}
\begin{align} 
  \psis &= \left[\frac{\partial W'(\is, \thetam)}{\partial \is}\right]^\T \\
  \taum &= \is^\T\J\psis + \frac{\partial W'(\is, \thetam)}{\partial \thetam} .
\end{align}
\end{subequations}
Note that the sign of the partial derivative term in the torque expression is positive. Similarly, the gradient relation in \eqref{eq:dual_mag} defines the flux-linkage map $\psis(\is, \thetam)$, which simplifies to $\psis(\is)$ without spatial harmonics. The choice between the energy-based model \eqref{eq:mag} and the co-energy-based model \eqref{eq:dual_mag} depends on whether flux linkage or current is chosen as the input in a given application.

\subsubsection{Rotor-Angle Representation} \label{sec:angle}
Due to the machine geometry, the magnetic characteristics repeat $k$ times per electrical revolution, i.e., they are periodic in the rotor angle with period $2\piup/k$. To represent this periodicity without discontinuities, the rotor angle is mapped to the unit vector
\begin{subequations} \label{eq:mag_per}
\begin{equation} \label{eq:encode_theta_fourier}
  \thetaper = \begin{bmatrix} \cos(k \thetam) \\ \sin(k \thetam) \end{bmatrix} .
\end{equation}
In machine learning, such encodings are often called Fourier features \cite{Tan2020}. Any function of $\thetam$ with period $2\piup/k$ can be expressed as a function of $\thetaper$, which thus also represents the harmonics of orders $2k$, $3k$, and so on. Expressing the field energy as $W(\psis,\thetam) = \Wper(\psis,\thetaper)$ and applying the chain rule with $\D\thetaper/\D\thetam = k\J\thetaper$, the model \eqref{eq:mag} becomes 
\begin{align}
  \is &= \left[\frac{\partial \Wper(\psis, \thetaper)}{\partial \psis}\right]^\T \\
  \tauper &= \left[\frac{\partial \Wper(\psis, \thetaper)}{\partial \thetaper}\right]^\T \\
  \taum &= \is^\T\J\psis + k \thetaper^\T \J \tauper
\end{align}
\end{subequations}
where the auxiliary vector $\tauper$ is the gradient of the energy function with respect to $\thetaper$. The analogous transformation applies to the co-energy-based model \eqref{eq:dual_mag}, where the sign of the last term in the torque expression becomes negative.

The model \eqref{eq:mag_per} uses the function $\Wper(\psis, \thetaper)$ only on the unit circle $\|\thetaper\| = 1$. Adding the term $c(\|\thetaper\|^2 - 1)$ to this function changes neither the current nor the torque, since the term does not depend on $\psis$ and $\thetaper^\T\J\thetaper = 0$. Since the incremental inverse inductance matrix is positive definite, a smooth energy function with this term is convex with respect to both $\psis$ and $\thetaper$ in a bounded operating range when the constant $c$ is sufficiently large. Therefore, the energy function can be assumed to be convex with respect to both its arguments without loss of generality.

\subsubsection{Symmetries From the Magnetic Geometry}
Assuming that the d-axis is aligned with the PM flux, the magnetic geometry is symmetric about the d-axis. Consequently, reversing the signs of the q-axis flux linkage and rotor angle negates the q-axis current but leaves the d-axis current unchanged \cite{Boj2026}
\begin{equation} \label{eq:is_symmetry}
  \is(\Conj\psis, \Conj \thetaper) = \Conj\is(\psis,\thetaper) \qquad
  \Conj = \mathrm{diag}(1,-1)
\end{equation}
where $\Conj$ negates the second component of a vector. In the following, the property \eqref{eq:is_symmetry} is referred to as q-axis symmetry. The symmetry of the geometry also implies that the field energy is an even function with respect to this reflection, $\Wper(\psis,\thetaper) = \Wper(\Conj\psis,\Conj\thetaper)$. To enforce the q-axis symmetry, the energy function is symmetrized as
\begin{equation} \label{eq:Wsym_def}
  \Wsym(\psis,\thetaper) = \frac{1}{2}\left[\Wper(\psis,\thetaper) + \Wper(\Conj\psis,\Conj\thetaper)\right] .
\end{equation}
Replacing $\Wper$ with $\Wsym$ in \eqref{eq:mag_per} guarantees the symmetry \eqref{eq:is_symmetry}. The same formulation applies to the co-energy function $\Wper'(\is, \thetaper)$. Without spatial harmonics, the angle dependence is dropped, and \eqref{eq:is_symmetry} reduces to the well-known relations $\isd(\psisd, -\psisq) = \isd(\psisd, \psisq)$ and $\isq(\psisd, -\psisq) = -\isq(\psisd, \psisq)$ \cite{Arm2013}. 

\begin{figure}[t] \centering
    \includegraphics{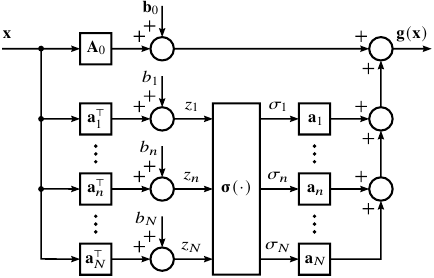} 
    \caption{Gradient network used for the magnetic models.} \label{fig:nn} 
\end{figure}

\section{Proposed Gradient-Network Magnetic Models} \label{sec:proposed_model}

\subsection{Gradient Networks} \label{sec:gradnet}
This subsection reviews the gradient networks of~\cite{Cha2025}. Furthermore, the $p$-norm gradient activation \eqref{eq:pnormgrad} and guidelines for selecting the activation functions are proposed. The magnetic models of synchronous machines based on these networks are developed in Section~\ref{sec:architectures}.

\subsubsection{Structure} 
Gradient networks are neural networks that are gradients of scalar functions by their structure \cite{Cha2025}. Furthermore, they can be parameterized to guarantee monotonicity, which is the relevant case for the proposed magnetic models. We use an architecture with $N$ hidden units, illustrated in Fig.~\ref{fig:nn}, given by
\begin{subequations} \label{eq:gradnet}
\begin{align} 
  \mathbf{z} &= \mathbf{A} \mathbf{x} + \mathbf{b} \\ 
  \g(\mathbf{x}) &= \mathbf{A}_0 \mathbf{x} + \mathbf{b}_0 + \mathbf{A}^\T \bm{\sigmaup}(\mathbf{z}) 
\end{align}
\end{subequations}
where $\mathbf{A} = [\mathbf{a}_1 \ldots \mathbf{a}_N]^\T$ is the weight matrix, $\mathbf{b} = [b_1 \ldots b_N]^\T$ is the bias vector, $\bm{\sigmaup}(\mathbf{z})$ is the activation function, $\mathbf{A}_0$ is the diagonal output matrix, and $\mathbf{b}_0$ is the output bias.

The network \eqref{eq:gradnet} is conservative for any $\mathbf{A}$, $\mathbf{b}$, $\mathbf{A}_0$, and $\mathbf{b}_0$: its Jacobian $\mathbf{J}_\g = \partial \g/\partial \mathbf{x}$ is symmetric whenever the activation Jacobian $\mathbf{J}_{\bm{\sigmaup}} = \partial \bm{\sigmaup}/\partial \mathbf{z}$ is symmetric (see Appendix~\ref{app:jacobian}). To also guarantee monotonicity, $\mathbf{A}_0$ and $\mathbf{J}_{\bm{\sigmaup}}$ are chosen to be positive semidefinite. For elementwise activations $\bm{\sigmaup}(\mathbf{z}) = [\sigma_1(z_1) \ldots \sigma_N(z_N)]^\T$, this simply requires each $\sigma_n$ to be non-decreasing. Consequently, $\g(\mathbf{x})$ represents the gradient of an underlying convex energy function $W(\mathbf{x})$, i.e., $\g(\mathbf{x}) = [\partial W(\mathbf{x})/\partial \mathbf{x}]^\T$, without the need to model or differentiate $W(\mathbf{x})$ explicitly.

\subsubsection{Elementwise Activations}
Elementwise activations apply a scalar function $\sigma(z)$ independently to each hidden unit. While computationally efficient, they represent only a subset of monotone gradient fields and lack universal approximation guarantees \cite{Cha2025}.

\begin{figure}[t] \centering
  \includegraphics[scale=.9]{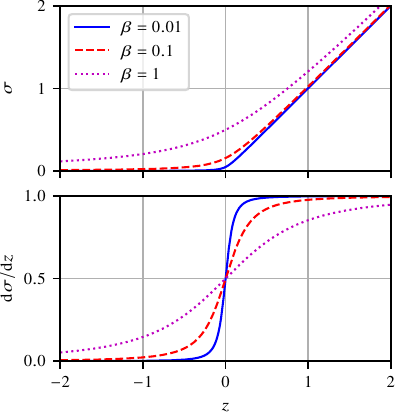}
  \caption{Elementwise squareplus activation $\sigma$ in \eqref{eq:squareplus} and its derivative $\D \sigma/\D z$ at different values of parameter $\beta$. The shape of the algebraic sigmoid \eqref{eq:sigmoid} is the same as the derivative of the squareplus, but shifted vertically and scaled.} \label{fig:activation}
\end{figure}

To capture magnetic saturation, the activation type must match the curvature of the modeled relationship. In the current map $\is(\psis)$, saturation causes the current to rise progressively with increasing flux linkage \cite{Fis1956}. This behavior is matched by rectifier-type activations with a non-decreasing derivative, such as the algebraic squareplus \cite{Bar2021}
\begin{equation} \label{eq:squareplus}
  \sigma(z) = \frac{1}{2}\left(z + \sqrt{z^2 + \beta}\right) 
\end{equation}
where the positive parameter $\beta$ tunes the smoothness around zero. As shown in Fig.~\ref{fig:activation}, its derivative $\D \sigma/\D z = (1 + z / \sqrt{z^2 + \beta})/2$ smoothly transitions from zero to one. Compared to the softplus, squareplus avoids exponential functions and is computationally more efficient.

Conversely, in the flux-linkage map $\psis(\is)$, the flux linkage saturates with increasing current, matching the familiar flattening of magnetic saturation curves. This behavior suits sigmoid-type activations. Here, we use the algebraic sigmoid
\begin{equation} \label{eq:sigmoid}   
  \sigma(z) = \frac{z}{\sqrt{z^2 + \beta}}
\end{equation}
which corresponds to a vertically shifted and scaled version of $\D\sigma/\D z$ in \eqref{eq:squareplus}.

\subsubsection{Vector Activations}
\begin{figure}[t] \centering
  \subfigure[]{
    \includegraphics[scale=.85]{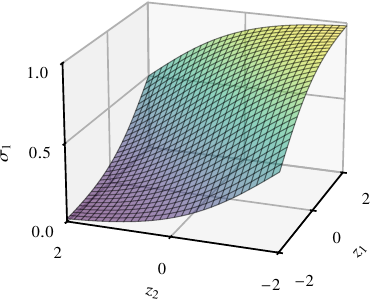}}
  \hfill
  \subfigure[]{
    \includegraphics[scale=.85]{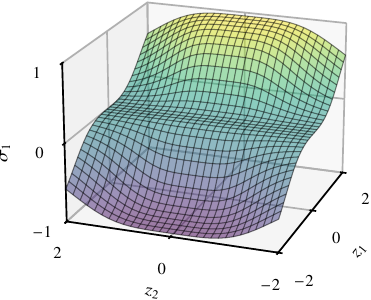}}
  \caption{Vector activation $\sigma_1(z_1, z_2)$ visualized in the two-dimensional case: (a) softmax \eqref{eq:softmax} with $\beta=1$; (b) $p$-norm gradient \eqref{eq:pnormgrad} with $p=4$ and $\beta=1$.} \label{fig:vector_act}
\end{figure}

Unlike elementwise activations, vector activations produce outputs where each component depends on the entire input vector $\mathbf{z}$. A standard vector activation is the scaled softmax, given by
\begin{align} \label{eq:softmax}
  \bm{\sigmaup}(\mathbf{z}) = \frac{1}{\sum_{n=1}^N \mathrm{e}^{\beta z_n}}\begin{bmatrix} \mathrm{e}^{\beta z_1} \\[-.25em] \vdots \\[.25em] \mathrm{e}^{\beta z_N} \end{bmatrix}
\end{align}
where the positive learnable parameter $\beta$ scales the inputs. Fig.~\ref{fig:vector_act}(a) illustrates the first component $\sigma_1(z_1, z_2)$ for the two-dimensional case. The softmax is the gradient of the convex log-sum-exp function $S(\mathbf{z}) = \log(\sum_{n=1}^N \mathrm{e}^{\beta z_n})/\beta$, which guarantees that its Jacobian $\mathbf{J}_{\bm{\sigmaup}}$ is symmetric and positive semidefinite. Furthermore, if $N$ and $\beta$ are sufficiently large, the network \eqref{eq:gradnet} with the softmax activation can approximate the gradient of any continuously differentiable convex function with an arbitrarily small maximum error in a given bounded range of the inputs \cite{Cha2025}. The term $\mathbf{A}_0\mathbf{x}$ does not affect this property, since the diagonal elements of $\mathbf{A}_0$ can be arbitrarily small.

As a computationally efficient alternative, we propose the $p$-norm gradient activation shown in Fig.~\ref{fig:vector_act}(b), defined as
\begin{align} \label{eq:pnormgrad}
  \bm{\sigmaup}(\mathbf{z}) = \frac{1}{\left[1 + \sum_{n=1}^N (\beta z_n)^{p}\right]^{\frac{p-1}{p}}} \begin{bmatrix} (\beta z_1)^{p-1} \\[-.25em] \vdots \\[.25em] (\beta z_N)^{p-1} \end{bmatrix}
\end{align}
where $p$ is a positive even integer and $\beta > 0$ is a learnable parameter. This activation is the gradient of the smooth regularized $p$-norm $S(\mathbf{z}) = [1 + \sum_{n=1}^N (\beta z_n)^{p}]^{(1/p)}/\beta$. The convexity of $S(\mathbf{z})$ guarantees monotonicity, while the unit term ensures differentiability at the origin and prevents division by zero. When $p$ is chosen as a power of two, transcendental functions are avoided completely: powers and fractional roots reduce to repeated multiplications and square roots, whereas the softmax requires $N$ exponential evaluations per point.\footnote{For example, for $p=8$, the fractional power in \eqref{eq:pnormgrad} is $T^{7/8} = (\sqrt{\sqrt{\sqrt{T}}})^7$, where $T = 1 + \sum_{n=1}^N (\beta z_n)^8$. The flux-linkage maps with the softmax \eqref{eq:softmax} and $p$-norm gradient \eqref{eq:pnormgrad} activations were compiled to C code and profiled on an Intel Core Ultra 9 285K CPU. Profiling indicates that the $p$-norm gradient executes over three times faster than the softmax across different network sizes. With per-unit inputs, the values of $|\beta z_n|$ remain small (below 8 in the $p$-norm gradient models of Table~\ref{tab:comparison_flux}), and the powers $(\beta z_n)^8$ cannot overflow even in single-precision arithmetic.} For small values of $|\beta z_n|$, the components of \eqref{eq:pnormgrad} are approximately the powers $(\beta z_n)^{p-1}$. In a current map, the choice $p=8$ thus resembles the power functions with empirical exponents used to describe iron saturation curves \cite{Fis1956, Qu2012, Hin2017}.

Both vector activations \eqref{eq:softmax} and \eqref{eq:pnormgrad} can be used in current and flux-linkage maps, with or without spatial harmonics. Only the softmax has the universal approximation property, but the proposed $p$-norm gradient gives comparable accuracy (Section~\ref{sec:results}) at a much lower computational cost.

\subsubsection{Behavior Outside the Data Range}
Outside the range of the training data, the behavior of the models depends on the activation function. The algebraic sigmoid, softmax, and $p$-norm gradient activations are bounded. Far outside the data range, the output of the network \eqref{eq:gradnet} then increases only through the term $\mathbf{A}_0\mathbf{x}$. For a flux-linkage map, this behavior corresponds to constant incremental inductances, which is physically expected at deep saturation. For a current map, however, the current would increase only linearly with the flux linkage, whereas the current of a saturating machine increases progressively. The squareplus activation is unbounded, and the hidden units then also contribute to the slope of the current map. Therefore, if the model has to be evaluated outside the range of the training data, the flux-linkage map with a bounded activation or the current map with the squareplus activation is preferred.

\begin{figure}[t] \centering
  \setlength{\subfigcapskip}{4pt}
  \subfigure[]{
    \includegraphics{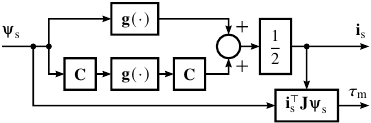}}
  \hfill
  \subfigure[]{
    \includegraphics{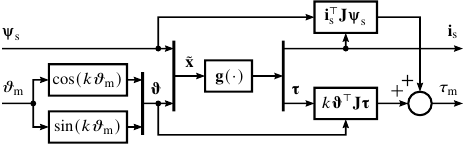}}
  \caption{Proposed magnetic models in rotor coordinates: (a) without spatial harmonics, including the symmetrization \eqref{eq:is_sym}; (b) with spatial harmonics, shown without the symmetrization for simplicity. In both cases, the function $\g(\cdot)$ is given by \eqref{eq:gradnet}.} \label{fig:prop_model}
\end{figure}

\subsection{Magnetic Model Architectures} \label{sec:architectures}
This subsection presents the proposed gradient-network architectures in rotor coordinates, using the energy-based current map as the primary example. The same formulation applies analogously to the co-energy-based flux-linkage map.

\subsubsection{Without Spatial Harmonics}
When spatial harmonics are omitted, the field energy depends only on the flux linkage. Without the symmetrized energy \eqref{eq:Wsym_def}, the current map can be parameterized directly as $\is = \g(\psis)$ using the monotone gradient network \eqref{eq:gradnet}. Setting $\mathbf{A}_0 = \mathrm{diag}(\mu_\mathrm{d}, \mu_\mathrm{q})$ enforces strong convexity with respect to the flux linkage. The parameters $\mu_\mathrm{d} > 0$ and $\mu_\mathrm{q} > 0$ provide lower bounds for the incremental inverse inductances and are learnable. Consequently, the Jacobian of the current map is positive definite, making the map strongly monotone \cite{Boy2004} and robustly invertible to obtain the flux linkage from the current.

With the symmetrized energy function \eqref{eq:Wsym_def}, the current map becomes
\begin{align} \label{eq:is_sym}
  \is(\psis) = \frac{1}{2}\left[ \g(\psis) + \Conj\,\g(\Conj\psis) \right]
\end{align}
which inherently satisfies the q-axis symmetry \eqref{eq:is_symmetry} while preserving the positive definiteness of the Jacobian and the resulting invertibility. Fig.~\ref{fig:prop_model}(a) shows the corresponding block diagram.

\subsubsection{With Spatial Harmonics}
Fig.~\ref{fig:prop_model}(b) shows the block diagram for the magnetic model with spatial harmonics, shown without the symmetrization step for simplicity. The combined input to the gradient network is $\xper = [\psis^\T, \thetaper^\T]^\T$, and its four outputs correspond to $\is$ and $\tauper$ in \eqref{eq:mag_per}. Strong monotonicity with respect to the flux linkage is enforced by setting $\mathbf{A}_0 = \mathrm{diag}(\mu_\mathrm{d}, \mu_\mathrm{q}, 0, 0)$. The monotone network makes the energy function convex with respect to both $\psis$ and $\thetaper$, which does not restrict the model (Section~\ref{sec:angle}). Hence, the universal approximation property of the softmax activation also holds with spatial harmonics.\footnote{Temperature could also be included as an input, with the corresponding output omitted. Since adding a quadratic term in temperature changes neither the current nor the torque, the convexity with respect to temperature does not restrict the model (cf.\ Section~\ref{sec:angle}). The symmetrization \eqref{eq:Wsym_def} keeps temperature unchanged.}

Analogously to \eqref{eq:is_sym}, the symmetrized energy $\Wsym(\psis,\thetaper)$ in \eqref{eq:Wsym_def} is implemented by symmetrizing the network outputs,
\begin{align} \label{eq:grad_sym_per}
  \begin{bmatrix} \is \\ \tauper \end{bmatrix} = \frac{1}{2}\left[ \g(\xper) + \tilde{\Conj}\,\g(\tilde{\Conj}\xper) \right]
\end{align}
where $\tilde{\Conj} = \mathrm{diag}(\Conj, \Conj)$. This symmetrization guarantees the q-axis symmetry \eqref{eq:is_symmetry}, while preserving strong monotonicity in $\psis$ for each $\thetam$ and periodicity in the rotor angle.

\begin{table}[t] \centering
\caption{Rated Values of the 5.6-kW PM Synchronous\\ Reluctance Machine} \label{tab:rating}
  \begin{tabular}{l | c | c}
  \toprule
  Voltage (line-to-neutral, peak value) & $\sqrt{2/3}\cdot 460$ V & 1.00 p.u. \\ 
  Current (peak value) & $\sqrt{2}\cdot 8.8$ A & 1.00 p.u. \\
  Frequency & 60 Hz & 1.00 p.u. \\
  Speed & 1\,800 r/min & 1.00 p.u. \\ 
  Power & 5.6 kW & 0.80 p.u. \\
  Torque & 29.7 Nm & 0.80 p.u. \\ \bottomrule
  \end{tabular}
\end{table}

\section{Results} \label{sec:results}
\begin{figure}[t] \centering
  \includegraphics[scale=.7]{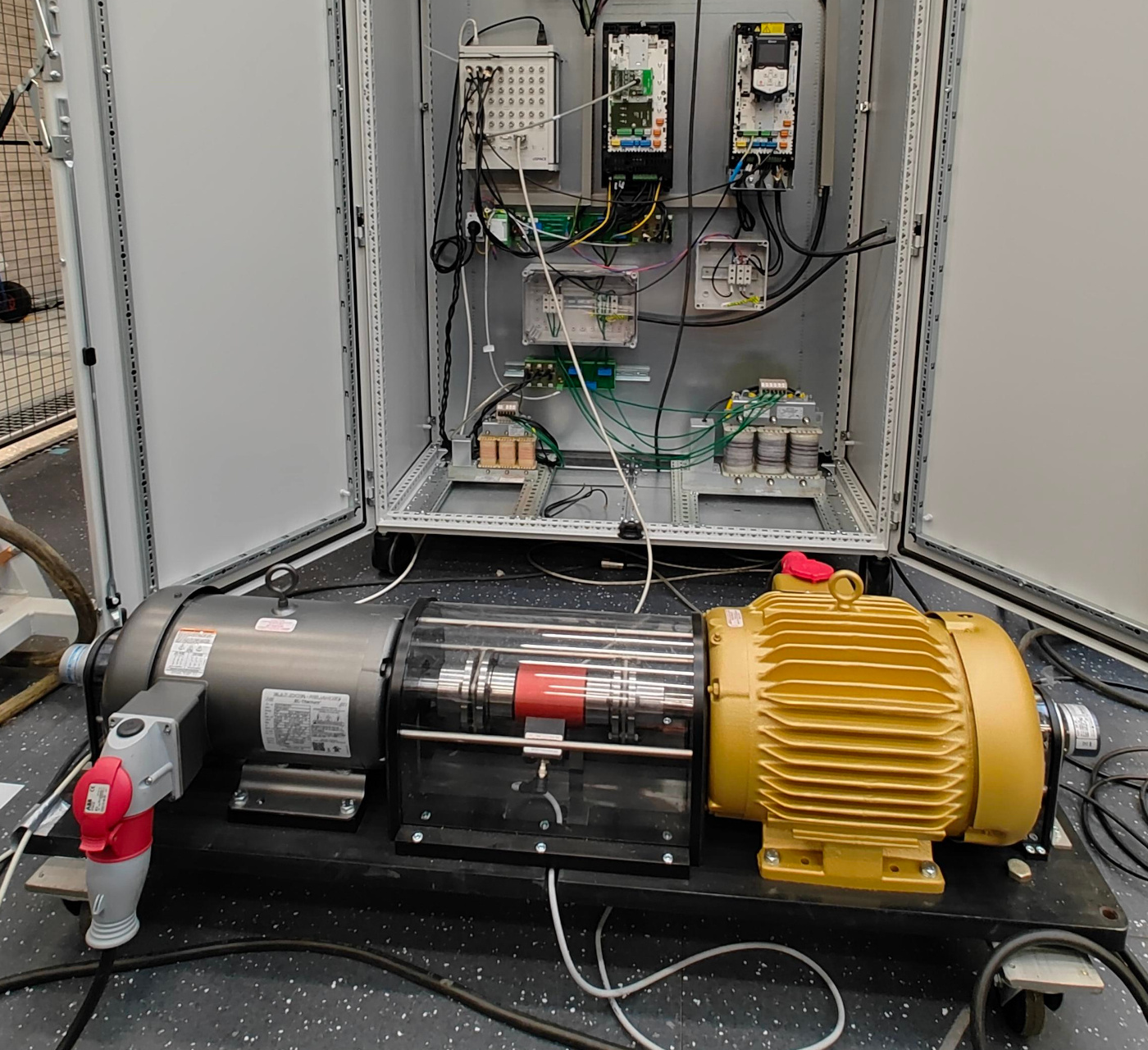}
  \caption{Test bench including a 5.6-kW PM synchronous reluctance machine (left), load machine (right), and their inverter cabinet (background).} \label{fig:baldor}
\end{figure}

Measured and FEM datasets from a four-pole 5.6-kW PM synchronous reluctance machine (ABB Baldor ECS101\,M0\,H7 EF4) were used to validate the proposed magnetic models. Table~\ref{tab:rating} lists the rated values of the machine, and Fig.~\ref{fig:baldor} shows the test bench used for measurements. Due to the rotor flux barriers, a small effective air gap in the q-axis direction, and the presence of PMs, the machine has highly nonlinear magnetic characteristics. Furthermore, a measured dataset from an automotive 10-pole 28-kW PM synchronous reluctance machine (Brusa HSM1-10.18.04) was used.\footnote{The datasets of both machines, the training code, and trained models are available in the open-source package \emph{motulator-gradnet} (version 0.2.0): \url{https://github.com/Aalto-Electric-Drives/motulator-gradnet}. Simulation examples, in which the trained models are used in both the machine model and the control system, are available in \emph{motulator} (version 0.26.2): \url{https://github.com/Aalto-Electric-Drives/motulator}.} Table~\ref{tab:models} summarizes the trained models. All models are trained offline. The model used in the control system is evaluated online in each control cycle.

The models were trained using small subsets of the datasets. A subset was obtained by taking every 10th or every 50th point of the dataset, which corresponds to 10\% or 2\% of the points, respectively. The remaining points were used for validation. To take into account the effect of the selected points, the training was repeated for every possible starting point of the selection, resulting in 10 runs for the 10\% subsets and 50 runs for the 2\% subsets. The tables give the mean and the standard deviation of the errors over these runs. For the measured dataset of the 5.6-kW machine, which has fewer points than the other datasets, a regular grid of 24 points was used instead of the 2\% subsets (Section~\ref{sec:results_meas}).

The training was implemented using the \textsl{PyTorch} library and the \textsl{AdamW} optimizer with a learning rate of 0.001, a weight decay of 0.01, and a batch size of 128. When the training subset had fewer points than the batch size, all its points were used in each training step. The number of epochs was fixed to 20\,000 for the measured datasets and 1\,000 for the FEM dataset, without early stopping. The weights were initialized from the normal distribution with zero mean and variance $2/(N + n)$, where $n$ is the number of inputs, using a different seed in each run. The biases were initialized to zero. The positive parameters $\mu_\mathrm{d}$, $\mu_\mathrm{q}$, and $\beta$ were parameterized as exponentials of learnable parameters with fixed initial values.

\begin{table*}[t] \centering
\caption{Summary of the Trained Models} \label{tab:models}
\begin{tabular}{l|l|l|l|l|l|c|c} \toprule
Section & Purpose & Inputs & Outputs & Training data & Activations & $N$ & Parameters \\ \midrule
\ref{sec:results_meas} & Current maps & $\psis$ & $\is$ & 5.6 kW, measured & squareplus, softmax, $p$-norm gradient & 12 & 41 \\
\ref{sec:results_meas} & Flux-linkage maps & $\is$ & $\psis$ & 5.6 kW, measured & algebraic sigmoid, softmax, $p$-norm gradient & 12 & 41 \\
\ref{sec:results_brusa} & Current map & $\psis$ & $\is$ & 28 kW, measured & squareplus & 12 & 41 \\
\ref{sec:results_brusa} & Flux-linkage map & $\is$ & $\psis$ & 28 kW, measured & $p$-norm gradient & 12 & 41 \\
\ref{sec:results_fem} & Spatial harmonics & $\is$, $\thetam$ & $\psis$, $\taum$ & 5.6 kW, FEM & softmax, $p$-norm gradient & 48 & 247 \\
\ref{sec:application} & Simulated machine & $\psis$, $\thetam$ & $\is$, $\taum$ & 5.6 kW, FEM & softmax & 48 & 247 \\
\ref{sec:application} & Control system & $\is$ & $\psis$ & 5.6 kW, measured & $p$-norm gradient & 6 & 23 \\ \bottomrule
\end{tabular}
\end{table*}

\begin{figure}[t] \centering
  \subfigure[]{\includegraphics[scale=.85]{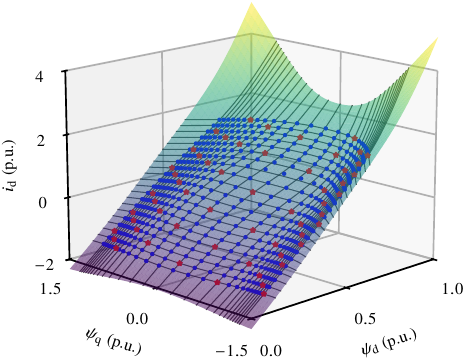}}
  \subfigure[]{\includegraphics[scale=.85]{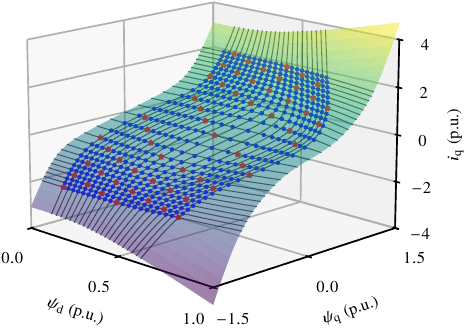}}
  \caption{Current maps of the 5.6-kW machine: (a) $\isd(\psisd, \psisq)$; (b) $\isq(\psisd, \psisq)$. The surfaces show the predicted maps from the model with the squareplus activation \eqref{eq:squareplus} and $N=12$ hidden units, trained on a 10\% subset. Markers show the measured dataset: red indicates the 10\% training subset and blue the remaining points. Gray lines show constant-current contours corresponding to the measured dataset.} \label{fig:meas_current_maps} 
\end{figure}

\begin{figure}[t] \centering
  \subfigure[]{\includegraphics[scale=.85]{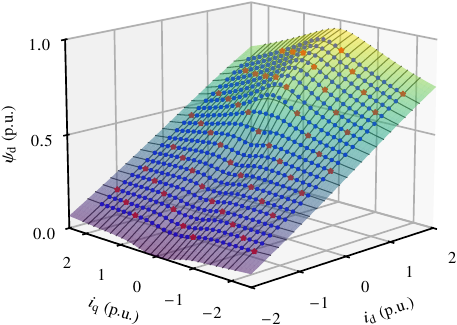}} 
  \subfigure[]{\includegraphics[scale=.85]{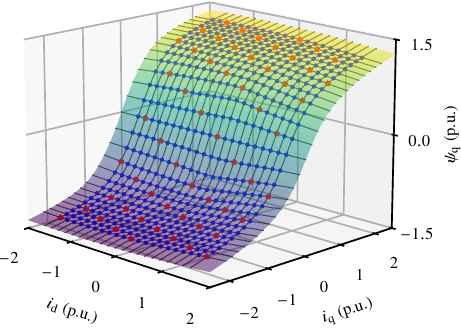}}
  \caption{Flux-linkage maps of the 5.6-kW machine: (a) $\psisd(\isd, \isq)$; (b) $\psisq(\isd, \isq)$. The surfaces show the predicted maps from the $p$-norm gradient model \eqref{eq:pnormgrad} with $p=8$ and $N=12$ hidden units, trained on a 10\% subset. Markers show the same measured dataset as in Fig.~\ref{fig:meas_current_maps}: red indicates the 10\% training subset and blue the remaining points. Gray lines show constant-current contours corresponding to the measured dataset.} \label{fig:meas_flux_maps} 
\end{figure}

\subsection{Measured Dataset of the 5.6-kW Machine} \label{sec:results_meas}

\subsubsection{Dataset}
The flux linkages of the 5.6-kW machine were measured on an equidistant current grid of $21 \times 27$ points using the constant-speed test \cite{Arm2013} at 400~r/min. The voltages were calculated from the measured DC-bus voltage and the switching states of the inverter. To reduce the measurement noise, the voltages were averaged over a steady-state interval at each operating point. The inverter nonlinearities (dead-time effect and threshold voltages of the power semiconductor devices) were characterized before the test and compensated. In this test, the resistive voltage drop is canceled by averaging the voltages measured in the motoring and regenerating modes. The test was also carried out at 600~r/min with practically the same results, which indicates that the effect of the core losses on the measured flux linkages is negligible.

According to the q-axis symmetry, the flux linkages at the currents $(\isd, \isq)$ and $(\isd, -\isq)$ are mirror images of each other. The flux linkages measured at these pairs of currents were averaged accordingly, and hence the dataset contains $21 \times 14$ unique points with nonnegative q-axis current. Figs.~\ref{fig:meas_current_maps} and~\ref{fig:meas_flux_maps} show this dataset in flux-linkage coordinates and current coordinates, respectively, including the mirrored points (567 points per map).

In addition to the 10\% subsets, the models were trained using a regular grid of $6 \times 4 = 24$ measured points, obtained by taking every fourth point of the measurement grid in both axes at positive q-axis currents. This grid covers the data range, including its boundary, and it corresponds to a measurement with a four times larger current step. The remaining points, excluding the mirror images of the training points, were used for validation. The training was repeated with 10 different initial weights.

\subsubsection{Model Configuration}
Both energy-based current maps $\is(\psis)$ and co-energy-based flux-linkage maps $\psis(\is)$ were trained using the proposed approach. The q-axis symmetry is enforced according to Fig.~\ref{fig:prop_model}(a). Different activation functions \eqref{eq:squareplus}--\eqref{eq:pnormgrad} are compared.

The magnetic models without spatial harmonics have two scalar inputs and two scalar outputs. The number of learnable parameters depends on the number $N$ of hidden units. The models have an $N \times 2$ weight matrix $\mathbf{A}$, an $N \times 1$ bias vector $\mathbf{b}$, a $2\times 2$ diagonal output matrix $\mathbf{A}_0$, a $2 \times 1$ output bias $\mathbf{b}_0$, and a single learnable activation parameter $\beta$. Consequently, the total number of learnable parameters is $3N + 5$. In the following, we use $N=12$, resulting in 41 learnable parameters. This number of parameters is small compared to the lookup-table approach.

\begin{table}[t] \centering
\caption{Comparison of Activation Functions in Current Maps, Trained Using Measured Data of the 5.6-kW Machine ($N=12$, Mean $\pm$ Standard Deviation Over the Runs)} \label{tab:comparison_curr}
\begin{tabular}{l|c|cc}
\toprule
Activation & Training data & $e_\mathrm{rms}$ (p.u.) & $e_\mathrm{max}$ (p.u.) \\ \midrule
\multirow{2}{*}{squareplus \eqref{eq:squareplus}} & 10\% & $0.014 \pm 0.007$ & $0.062 \pm 0.031$ \\ 
                                                  & 24 points & $0.019 \pm 0.003$ & $0.077 \pm 0.008$ \\ \midrule
\multirow{2}{*}{softmax \eqref{eq:softmax}}       & 10\% & $0.019 \pm 0.002$ & $0.073 \pm 0.015$ \\
                                                  & 24 points & $0.025 \pm 0.003$ & $0.095 \pm 0.010$ \\ \midrule
$p$-norm                                          & 10\% & $0.019 \pm 0.002$ & $0.079 \pm 0.022$ \\
gradient \eqref{eq:pnormgrad}                     & 24 points & $0.023 \pm 0.002$ & $0.091 \pm 0.007$ \\ \bottomrule
\end{tabular}
\end{table}

The mean squared error (MSE) loss was used for training current maps, defined as
\begin{equation} \label{eq:mse_loss_current}
  \mathcal{L} = \frac{1}{2L} \sum_{\ell=1}^{L} \big\| \isk - \hatisk \big\|^2
\end{equation}
where $L$ is the number of training samples, $\isk$ is the measured current for sample $\ell$, and $\hatisk(\psisk)$ is the model prediction for the corresponding flux-linkage input $\psisk$. For training flux-linkage maps, the same procedure was used, replacing the current error with the flux-linkage error.

\subsubsection{Model Performance}
Fig.~\ref{fig:meas_current_maps} visualizes the learned current maps with the squareplus activation~\eqref{eq:squareplus}, when every tenth data point of the full measured dataset is used for training and the rest for validation. Fig.~\ref{fig:meas_flux_maps} shows flux-linkage maps learned by training the co-energy-based model with the $p$-norm gradient activation~\eqref{eq:pnormgrad} and the same training dataset. To facilitate visual comparison, constant-current contours corresponding to the measurement grid are drawn on the predicted surfaces. Ideally, the measured points would fall exactly at the intersections of these contours. The measured points lie close to these intersections. The other activation functions yield visually similar results when trained with 10\% of the dataset.

For quantitative comparison, the root-mean-square (rms) error at the validation points is computed as
\begin{equation}
  e_\mathrm{rms} = \sqrt{\frac{1}{M} \sum_{\ell=1}^M e_\ell^2}
\end{equation}
where $M$ is the number of these points and $e_\ell = \|\isk - \hatisk \|$. Furthermore, the maximum error $e_\mathrm{max} = \max_\ell e_\ell$ is considered. The above metrics are defined for current maps, but they are similarly defined for flux-linkage maps. 

Table~\ref{tab:comparison_curr} gives the errors of the current maps for different activation functions. With 10\% of the dataset (about 57 points), the rms error is below 0.02~p.u.\ with all the activation functions, i.e., the elementwise squareplus activation is as accurate as the vector activations. With the 24-point grid, the errors are only slightly higher, even though the number of training points is less than half of that of the 10\% subsets.

Table~\ref{tab:comparison_flux} gives the results for flux-linkage maps. The models with vector activations show better accuracy than the model with the elementwise algebraic sigmoid, both with 10\% of the dataset and with the 24-point grid. Hence, the vector activations are preferable in flux-linkage maps.

\begin{table}[t] \centering
\caption{Comparison of Activation Functions in Flux-Linkage Maps, Trained Using Measured Data of the 5.6-kW Machine ($N=12$, Mean $\pm$ Standard Deviation Over the Runs)} \label{tab:comparison_flux}
\begin{tabular}{l|c|cc}
\toprule
Activation & Training data & $e_\mathrm{rms}$ (p.u.) & $e_\mathrm{max}$ (p.u.) \\ \midrule
algebraic                                   & 10\% & $0.012 \pm 0.004$ & $0.038 \pm 0.008$ \\ 
sigmoid \eqref{eq:sigmoid}                  & 24 points & $0.019 \pm 0.004$ & $0.056 \pm 0.012$ \\ \midrule
\multirow{2}{*}{softmax} & 10\% & $0.005 \pm 0.001$ & $0.026 \pm 0.007$ \\
                                            & 24 points & $0.007 \pm 0.002$ & $0.031 \pm 0.003$ \\ \midrule
$p$-norm                                    & 10\% & $0.007 \pm 0.002$ & $0.028 \pm 0.005$ \\
gradient              & 24 points & $0.009 \pm 0.002$ & $0.033 \pm 0.005$ \\ \bottomrule
\end{tabular}
\end{table}

The effect of the number of hidden units was studied by training the flux-linkage map with the $p$-norm gradient activation and 10\% of the dataset. With $N = 3$, 6, 12, and 24, the rms errors are 0.058, 0.011, 0.007, and 0.004~p.u., respectively. Hence, $N = 12$ is a reasonable compromise between accuracy and computational cost. Similarly, with $p = 4$, 8, and 16 ($N = 12$), the rms errors are 0.007, 0.007, and 0.017~p.u., respectively, i.e., the accuracy is the same with $p = 4$ and $p = 8$, but it decreases with $p = 16$, for which the training loss also decreases more slowly.

The placement of the training points is important. When a similar number of points is taken as every 25th point of the dataset, the rms errors of the squareplus current map and the $p$-norm gradient flux-linkage map increase to 0.048 and 0.022~p.u., respectively, i.e., more than twice as high as with the 24-point grid. The maximum errors then typically occur at the boundary of the data range. Therefore, the training points should cover the whole operating range, including its boundary.

Table~\ref{tab:gn_mlp_comparison} compares the gradient network ($p$-norm gradient activation) with other models of the flux-linkage map. All models were identified using the 24-point grid and evaluated at the same validation points. The lookup table consists of the 24 measured points and their mirror images, with bilinear interpolation. The analytical model \cite{Lel2024} is a current map developed for PM synchronous reluctance machines. Its 14 parameters were fitted using the nonlinear least-squares method, and the map was inverted numerically. The unconstrained multilayer perceptrons (MLPs) have one hidden layer of eight units. The Hamiltonian neural network (HNN)~\cite{Gre2019, Des2021} is a scalar co-energy network with one hidden layer of ten units, differentiated by automatic differentiation. The MLPs and the HNN use the same symmetrization as in \eqref{eq:is_sym} and the same training settings as the gradient network. Hence, all compared models satisfy the q-axis symmetry, i.e., the MLPs are not purely black-box models.

Reciprocity and monotonicity in Table~\ref{tab:gn_mlp_comparison} are evaluated from the incremental inductance matrix $\partial\psis/\partial\is$. The reciprocity residual is the mean of $|\partial\psisd/\partial\isq - \partial\psisq/\partial\isd|$ at the validation points. For comparison, the reciprocity residual of the measured data themselves is 0.002~p.u., whereas the mean of $|\partial\psisd/\partial\isq|$ in the data is 0.05~p.u.\ (both computed by means of central differences). A point is non-monotone if the symmetric part of $\partial\psis/\partial\is$ is not positive definite at this point. At the validation points, only two of the MLPs have such points (less than 1\% of the points). Table~\ref{tab:gn_mlp_comparison} therefore gives the share of the non-monotone points in a current range twice as wide as the data range in both axes, where the lookup table is not defined.

In the data range, the analytical model, the HNN, and the MLPs with the tanh and softmax activations are as accurate as the gradient network. The rms error of the lookup table is about five times higher. The analytical model is reciprocal and monotone, but its functional form is specific to the machine type and does not include spatial harmonics. The MLPs and the lookup table are not reciprocal. The HNN is reciprocal, but it is non-monotone at about half of the points of the wider range, i.e., its invertibility cannot be relied on outside the data range. The gradient network is reciprocal and monotone for any input.

\begin{table}[t] \centering
\caption{Comparison of Models of the Flux-Linkage Map of the 5.6-kW Machine, Identified Using the 24-Point Grid ($e_\mathrm{rms}$: Mean $\pm$ Standard Deviation Over the Runs)} \label{tab:gn_mlp_comparison}
\begin{tabular}{@{}l|c|c|c|c@{}} \toprule
\multirow{3}{*}{Model} & \multirow{3}{*}{\shortstack{Para-\\meters}} & \multirow{3}{*}{$e_\mathrm{rms}$ (p.u.)} & Reciprocity & Non- \\
 & & & residual & monotone \\
 & & & (p.u.) & points (\%) \\ \midrule
Gradient network & 41 & $0.009 \pm 0.002$ & 0 & 0 \\
Analytical \cite{Lel2024} & 14 & 0.010 & 0 & 0 \\
HNN (tanh) & 40 & $0.009 \pm 0.001$ & 0 & 52 \\
MLP (tanh) & 42 & $0.008 \pm 0.001$ & 0.004 & 0.7 \\
MLP (softmax) & 43 & $0.010 \pm 0.001$ & 0.009 & 12 \\
MLP ($p$-norm) & 43 & $0.025 \pm 0.012$ & 0.043 & 9 \\
Lookup table & 48 & 0.043 & 0.011 & -- \\ \bottomrule
\end{tabular}
\end{table}

\begin{figure}[t] \centering
  \subfigure[]{\includegraphics[scale=.85]{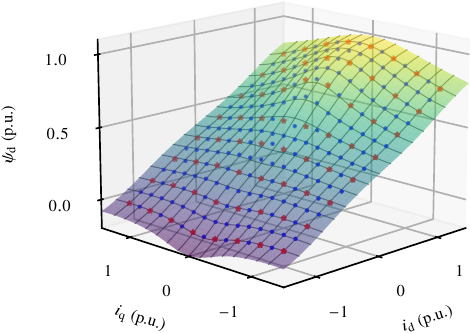}}
  \subfigure[]{\includegraphics[scale=.85]{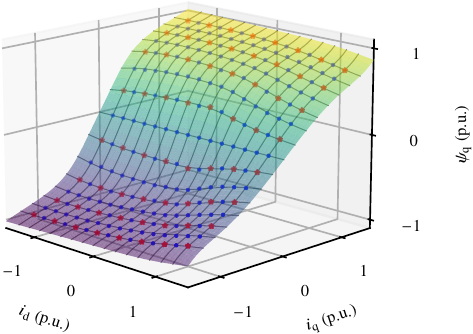}}
  \caption{Flux-linkage maps of the 28-kW machine: (a) $\psisd(\isd, \isq)$; (b) $\psisq(\isd, \isq)$. The surfaces show the predicted maps from the $p$-norm gradient model \eqref{eq:pnormgrad} with $p=8$ and $N=12$ hidden units, trained using the measured points on a grid with a step of 40~A (red markers). For clarity, blue markers show only every fourth measured point in both directions. Gray lines show the constant-current contours of the predicted maps at the currents of the markers.} \label{fig:brusa_flux_maps}
\end{figure}

\begin{table}[t] \centering
\caption{Errors of the Models of the 28-kW Machine ($N=12$, Mean $\pm$ Standard Deviation Over the Runs)} \label{tab:brusa}
\begin{tabular}{l|c|cc}
\toprule
Model & Training data & $e_\mathrm{rms}$ (p.u.) & $e_\mathrm{max}$ (p.u.) \\ \midrule
Flux-linkage map,           & 10\% & $0.009 \pm 0.002$ & $0.061 \pm 0.004$ \\
$p$-norm gradient           & 2\%  & $0.009 \pm 0.001$ & $0.062 \pm 0.006$ \\ \midrule
Current map,                & 10\% & $0.019 \pm 0.002$ & $0.136 \pm 0.012$ \\
squareplus                  & 2\%  & $0.019 \pm 0.002$ & $0.138 \pm 0.014$ \\ \bottomrule
\end{tabular}
\end{table}

\subsection{Measured Dataset of the 28-kW Machine} \label{sec:results_brusa}
The proposed approach was also applied to the 10-pole 28-kW machine. Its rated current is 78~A, and its rated torque is 52~Nm. The base values of the voltage, current, and frequency are $\sqrt{2/3}\cdot 283$~V, $\sqrt{2}\cdot 78$~A, and 383~Hz, respectively. The flux linkages were measured using the constant-speed test \cite{Arm2013} at 500~r/min \cite{Foi2026}. The torque was measured using a torque transducer. The current grid is equidistant with a step of 5~A, ranging from $-145$ to $145$~A ($\pm 1.31$~p.u.) in both axes. With the mirrored points, the dataset contains $59 \times 59 = 3\,481$ points. The model configuration and the training procedure are the same as in Section~\ref{sec:results_meas}, i.e., the models have $N=12$ hidden units and 41 learnable parameters.

Fig.~\ref{fig:brusa_flux_maps} shows the flux-linkage maps learned by the co-energy-based model with the $p$-norm gradient activation \eqref{eq:pnormgrad}. This model was trained using $8 \times 4 = 32$ measured points on a regular grid with a step of 40~A (64 points with their mirror images), which corresponds to a measurement with an eight times larger current step. Its rms error is 0.011~p.u. Table~\ref{tab:brusa} gives the errors of this model type and of the energy-based model with the squareplus activation \eqref{eq:squareplus}, when 10\% or 2\% of the points are used for training. With 10\% of the points, the rms error is 0.009~p.u.\ for the flux-linkage map and 0.019~p.u.\ for the current map. The errors are practically the same with 2\% of the points (about 70 points), since the measurement grid of this machine is dense. The maximum errors occur at low d-axis currents, where the measured d-axis flux linkage decreases with increasing d-axis current at a few grid points. A monotone model cannot reproduce such points, which are probably caused by measurement errors at low currents. With the same model size, the errors are of the same order as those of the 5.6-kW machine.

The dataset also includes the measured torque. The torque calculated from the flux-linkage map of Fig.~\ref{fig:brusa_flux_maps} differs from the measured torque by 0.81~Nm (rms value), which is 1.6\% of the rated torque. For comparison, the torque calculated directly from the measured flux linkages differs from the measured torque by 0.65~Nm.

\begin{figure}[t] \centering
  \subfigure[]{\includegraphics[scale=.85]{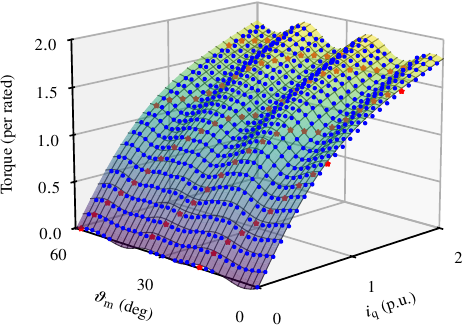}} 
  \subfigure[]{\includegraphics[scale=.85]{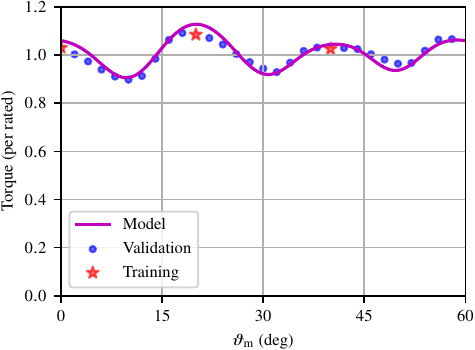}}
  \caption{Electromagnetic torque of the 5.6-kW machine from the $p$-norm gradient model with $N=48$ hidden units: (a) $\taum(\isq, \thetam)$ at constant $\isd$ corresponding to the rated MTPA current; (b) $\taum(\thetam)$ at the constant rated MTPA stator current. The curve and markers in (b) are a slice from (a). Red and blue markers show the full FEM dataset. Red markers indicate the 10\% training subset.} \label{fig:fem_torque_maps}
\end{figure}

\subsection{FEM Dataset With Spatial Harmonics} \label{sec:results_fem}

\subsubsection{Dataset}
To demonstrate the capability of the proposed model in Fig.~\ref{fig:prop_model}(b) in capturing spatial harmonics, the 5.6-kW machine was analyzed using FEM under magnetostatic conditions. For this machine, both the flux linkage and torque are periodic in the electrical angle $\thetam$ with a 60$^\circ$ period, i.e., $k = 6$ in \eqref{eq:encode_theta_fourier}. The FEM dataset was generated on an equidistant current grid ($61 \times 61$), ranging from $-2.41$ to $2.41$~p.u., and at 30 equidistant rotor angles from $0^\circ$ to $58^\circ$. The total dataset size is $61 \times 61 \times 30 = 111\,630$ samples. At each operating point, the d- and q-axis flux linkages and the electromagnetic torque were computed. The torque was computed using Arkkio's method \cite{Ark1987}, which is a variant of the Maxwell stress tensor method.

\subsubsection{Model Configuration}
Both the energy-based and co-energy-based models were trained using the proposed approach, with q-axis symmetry enforced via \eqref{eq:grad_sym_per}. For brevity, the results are shown only for the co-energy-based model, but the energy-based model shows similar performance. The vector activations softmax \eqref{eq:softmax} and the $p$-norm gradient \eqref{eq:pnormgrad} are compared. The training dataset size is either 10\% or 2\% of the full FEM dataset, corresponding to 11\,163 and 2\,232 samples, respectively.

The magnetic models with spatial harmonics have four scalar inputs and four scalar outputs.
The models have an $N \times 4$ weight matrix $\mathbf{A}$, an $N \times 1$ bias vector $\mathbf{b}$, two learnable diagonal elements in the matrix $\mathbf{A}_0$, a $4 \times 1$ output bias $\mathbf{b}_0$, and a single learnable activation parameter $\beta$. Consequently, the total number of learnable parameters is $5N + 7$. In the following examples, $N = 48$ hidden units are used, resulting in $247$ learnable parameters. More hidden units are needed than in Section~\ref{sec:results_meas}, since the model has two additional inputs and also has to reproduce the dependence on the rotor angle. The MSE loss is used for training, but it includes both the flux-linkage and torque errors. In the case of the co-energy-based models, the loss function is defined as
\begin{equation}
  \mathcal{L} = \frac{1}{L} \sum_{\ell=1}^{L}
  \left[ \frac{1}{2}\bigl\| \psisk - \hatpsisk \bigr\|^2 + \bigl(\tau_{\vartheta\ell} - \hat{\tau}_{\vartheta\ell}\bigr)^2\right]
\end{equation}
where $\tau_\vartheta = \taum - \is^\T\J\psis$ is the last term of the torque expression in \eqref{eq:mag_per}, i.e., the torque component obtained from the auxiliary vector. For sample $\ell$, the value $\tau_{\vartheta\ell}$ is calculated from the data and its prediction $\hat{\tau}_{\vartheta\ell}$ from the outputs of the model. When training energy-based models, the loss function is defined similarly, replacing the flux-linkage error term with the current error term.

\begin{table}[t] \centering
\caption{Comparison of Activation Functions in Models With Spatial Harmonics, Trained Using FEM Data of the 5.6-kW Machine ($N=48$, Mean $\pm$ Standard Deviation Over the Runs)} \label{tab:activation_comparison_harm}
\begin{tabular}{l|l|c|cc} \toprule
\multirow{2}{*}{Quantity} & \multirow{2}{*}{Activation} & Training & \multirow{2}{*}{$e_\mathrm{rms}$} & \multirow{2}{*}{$e_\mathrm{max}$} \\
&  & data  &  &       \\ \midrule
\multirow{4}{*}{\shortstack{Flux\\linkage\\(p.u.)}}
  & \multirow{2}{*}{softmax}  & 10\% & $0.009 \pm 0.000$ & $0.033 \pm 0.002$ \\
  &                           & 2\%  & $0.014 \pm 0.002$ & $0.053 \pm 0.007$ \\ \cmidrule(l){2-5}
  &$p$-norm                   & 10\% & $0.013 \pm 0.001$ & $0.049 \pm 0.005$ \\
  &gradient                   & 2\%  & $0.019 \pm 0.003$ & $0.069 \pm 0.011$ \\ \midrule
\multirow{4}{*}{\shortstack{Torque\\(per rated)}}
  & \multirow{2}{*}{softmax}  & 10\% & $0.023 \pm 0.001$ & $0.165 \pm 0.016$ \\
  &                           & 2\%  & $0.030 \pm 0.003$ & $0.204 \pm 0.026$ \\ \cmidrule(l){2-5}
  &$p$-norm                   & 10\% & $0.029 \pm 0.001$ & $0.222 \pm 0.019$ \\
  &gradient                   & 2\%  & $0.037 \pm 0.003$ & $0.261 \pm 0.024$ \\ \bottomrule
\end{tabular}
\end{table}

\begin{figure*}[t] \centering
    \subfigure[]{\includegraphics[scale=0.85]{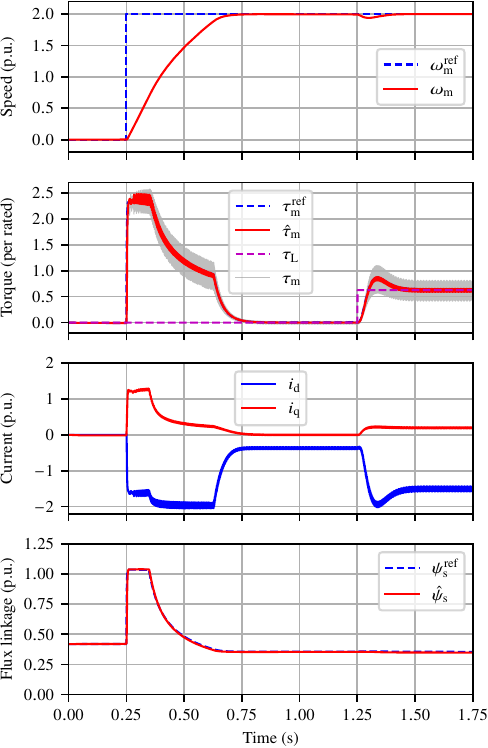}} \qquad\qquad
    \subfigure[]{\includegraphics[scale=0.85]{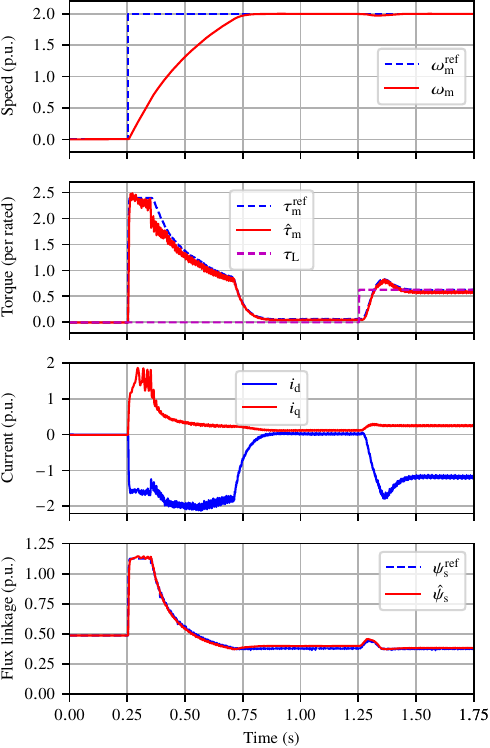}}
    \caption{Closed-loop control responses of the 5.6-kW machine: (a) simulation with the proposed magnetic model including spatial harmonics in the plant; (b) corresponding measurement. The same controller is used in both cases, parameterized entirely by the proposed model without spatial harmonics. Speed reference step from 0 to 2~p.u. at $t = 0.25$~s and load-torque step at $t = 1.25$~s. From top to bottom, the responses are speed, torque, current, and flux linkage. The measured torque is unavailable in (b).} \label{fig:sim_meas_res}
\end{figure*}

\subsubsection{Model Performance}
Fig.~\ref{fig:fem_torque_maps}(a) visualizes the proposed co-energy-based model with the $p$-norm gradient activation \eqref{eq:pnormgrad}, trained with 10\% of the FEM dataset. The model captures both the rotor-angle dependence and the q-axis current dependence of the torque. Fig.~\ref{fig:fem_torque_maps}(b) shows a slice at $\isd = -0.72$~p.u. and $\isq = 0.72$~p.u., which approximately corresponds to the rated maximum-torque-per-ampere (MTPA) current. The torque ripple caused by spatial harmonics is well captured.

Table~\ref{tab:activation_comparison_harm} summarizes the flux-linkage and torque prediction errors of the co-energy-based model at the validation points. The errors remain small even when the training dataset is limited to 2\% of the full FEM dataset. The $p$-norm gradient activation \eqref{eq:pnormgrad} yields slightly higher flux-linkage errors than the softmax activation \eqref{eq:softmax}. The errors are slightly higher with the 2\% training dataset than with the 10\% training dataset. The torque errors follow the same trends.

The effect of the number of hidden units was studied using the $p$-norm gradient activation and 10\% of the dataset. With $N = 12$, 24, 48, and 96, the rms flux-linkage errors are 0.031, 0.015, 0.013, and 0.012~p.u., respectively. Hence, $N = 48$ is a reasonable compromise between accuracy and computational cost.

\subsection{Application Examples} \label{sec:application}
The trained magnetic models are next evaluated in closed-loop simulations and experiments. Fig.~\ref{fig:baldor} shows the experimental setup, where the same 5.6-kW PM synchronous reluctance machine is controlled by a dSPACE MicroLabBox.

The controller is a flux-vector control scheme \cite{Tii2025,Var2022} parameterized by the proposed co-energy-based model \eqref{eq:dual_mag} without spatial harmonics [$p$-norm gradient activation \eqref{eq:pnormgrad} with $p=8$, $N=6$]. This model has 23 parameters. For comparison, lookup tables of the same maps on the measurement grid of $21 \times 27$ points would have 1\,134 elements, or 588 elements if the symmetry is exploited. The model was trained with 10\% of the measured data. The rms error of such models is 0.011~p.u.\ on average (Section~\ref{sec:results_meas}).

The proposed model is used in the control law, flux observer, and online reference generation. These control methods also require the incremental inductance matrix, i.e., the Jacobian of the flux-linkage map. Since the model is smooth, its Jacobian is continuous and can be computed in closed form (Appendix~\ref{app:jacobian}), whereas a linearly interpolated lookup table has discontinuous derivatives. The model and its Jacobian are evaluated at four operating points during each 8-kHz control cycle with single-update PWM. The applied online reference generation method \cite{Sar2026} numerically inverts the flux-linkage map to obtain the current for the maximum-torque-per-volt (MTPV) calculation, which requires the magnetic model to be physically consistent and invertible.

\subsubsection{Simulation}
In the simulation, the machine is represented by the proposed energy-based magnetic model \eqref{eq:mag_per} with spatial harmonics (softmax activation, $N=48$), trained on the FEM dataset as described in Section~\ref{sec:results_fem}. This model variant is convenient since the flux linkages are the state variables of the voltage equations \eqref{eq:dyn}. Fig.~\ref{fig:sim_meas_res}(a) shows the corresponding response. In the shown sequence, the drive operates in the MTPA, field-weakening, current-limit, and MTPV regions. Spatial-harmonic effects (the sixth electrical harmonic and its multiples) are most visible in the actual torque, but also appear in other signals.

\subsubsection{Experiment}
Fig.~\ref{fig:sim_meas_res}(b) shows the corresponding measurement using the same control method as in the simulation. In addition to the expected sixth electrical harmonic and its multiples, the measured q-axis current contains first- and fourth-order mechanical harmonics. These low-order components may be caused by practical asymmetries (e.g., eccentricity, ovality, or winding imbalance) not present in the symmetric FEM model. If necessary, such phenomena could be incorporated into the proposed model by introducing corresponding physical variables, such as lateral rotor displacement, as additional inputs.

Additionally, transients appear in the flux reference during field weakening due to actual DC-bus voltage variations, whereas the simulation assumed a constant DC-bus voltage. Otherwise, the simulation and measurement agree well (although the FEM dataset naturally only approximates the real machine because of uncertainties in geometry and material properties).

The experiment confirms that the proposed gradient-network model can be executed as an integral part of the real-time control system. At the 8-kHz sampling frequency (125-$\upmu$s sampling period), the average turnaround time of the control cycle is 73~$\upmu$s with the $p$-norm gradient and 105~$\upmu$s with the softmax, corresponding to 58\% and 84\% of the sampling period, respectively.

\section{Conclusions} \label{sec:conclusions}
A magnetic modeling framework for synchronous machines was presented. The current and the electromagnetic torque are modeled directly as the gradient of the field energy by means of a monotone gradient network (or, in the dual formulation, the flux linkage and torque as the gradient of the co-energy). Hence, the model inherently satisfies energy conservation and reciprocity, and the relationship between the flux linkages and currents is invertible. Spatial harmonics are included by means of a periodic representation of the rotor angle, and the q-axis symmetry is enforced by symmetrizing the energy function.

A model with a few tens of parameters, trained using a small number of measured points, reproduces the measured current and flux-linkage maps of two PM synchronous reluctance machines with small errors. With spatial harmonics, the model also reproduces the torque ripple obtained from FEM. The proposed $p$-norm gradient activation is computationally more efficient than the softmax activation. With this activation, the model was used in the control law, flux observer, and reference generation of a real-time control system.

\appendices

\section{Transformation to Rotor Coordinates} \label{app:transformation}
The same field energy can be expressed in different coordinates, $W(\psis, \thetam) = \Ws(\psiss, \thetam)$. The partial derivatives of $W$ are taken with $\psis$ and $\thetam$ as the independent variables, and those of $\Ws$ with $\psiss$ and $\thetam$. Using the chain rule with $\psiss = \e^{\thetam\J}\psis$ gives
\begin{subequations}
\begin{align}
  \frac{\partial W(\psis,\thetam)}{\partial \psis}
  &= \frac{\partial \Ws(\psiss,\thetam)}{\partial \psiss}\frac{\partial \psiss}{\partial \psis} = (\iss)^\T \e^{\thetam\J} = \is^\T \\ \nonumber
  \frac{\partial W(\psis,\thetam)}{\partial \thetam}
  &= \frac{\partial \Ws(\psiss,\thetam)}{\partial \psiss} \frac{\partial \psiss}{\partial \thetam}  
 + \frac{\partial \Ws(\psiss,\thetam)}{\partial \thetam} \\
  &= \is^\T \J\psis + \frac{\partial \Ws(\psiss,\thetam)}{\partial \thetam} .
\end{align}
\end{subequations}
Using $\taum = -\partial \Ws/\partial \thetam$ yields the expressions in \eqref{eq:mag}.

\section{Jacobian of the Gradient Network Model} \label{app:jacobian}
The vector field \eqref{eq:gradnet} is conservative (i.e., a gradient field) if its Jacobian $\mathbf{J}_\g$ is symmetric \cite{Cha2025}. Applying the chain rule, the Jacobian of the gradient network \eqref{eq:gradnet} becomes
\begin{align} 
  \mathbf{J}_\g(\mathbf{x}) = \frac{\partial \g(\mathbf{x})}{\partial \mathbf{x}}
  = \mathbf{A}_0 + \mathbf{A}^\T \frac{\partial \bm{\sigmaup}(\mathbf{z})}{\partial \mathbf{z}} \mathbf{A} .
\end{align}
The first term $\mathbf{A}_0$ is diagonal and thus symmetric. The second term is symmetric if $\mathbf{J}_{\bm{\sigmaup}} = \partial \bm{\sigmaup}/\partial \mathbf{z}$ is symmetric, which is the case when the activation $\bm{\sigmaup}$ is the gradient of a scalar function. Note that $\mathbf{J}_{\bm{\sigmaup}}$ is diagonal and thus symmetric when the activation is elementwise. The network \eqref{eq:gradnet} is monotone if $\mathbf{J}_\g$ is positive semidefinite \cite{Cha2025, Boy2004}. This condition holds if $\mathbf{A}_0$ and $\mathbf{J}_{\bm{\sigmaup}}$ are positive semidefinite.

\section*{Acknowledgment}
An artificial intelligence tool (Claude, Anthropic) was used to assist in reproducing the numerical results of Section~\ref{sec:results} and in editing the language.


\makeatletter\def\@IEEEBIOskipN{2\baselineskip}\makeatother 

\begin{IEEEbiography}[{\includegraphics[width=1in,height=1.25in,clip,keepaspectratio]{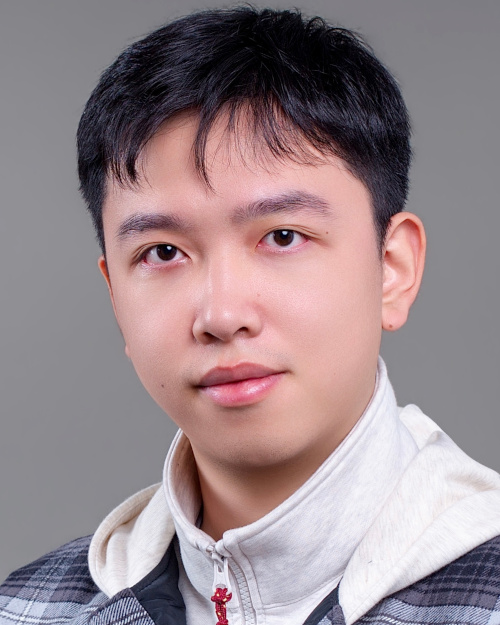}}]
{Junyi Li} (Graduate Student Member, IEEE) received the B.Sc. degree in communication engineering from the Taiyuan University of Technology, Taiyuan, China, in 2022, and the M.Sc. (Tech.) degree in information and communication engineering from Aalto University, Espoo, Finland, in 2025.  

He is currently working toward the D.Sc. (Tech.) degree in electrical engineering with Aalto University. His research interests include physics-informed machine learning and electric machine modeling for monitoring and control.
\end{IEEEbiography}

\begin{IEEEbiography}[{\includegraphics[width=1in,height=1.25in,clip,keepaspectratio]{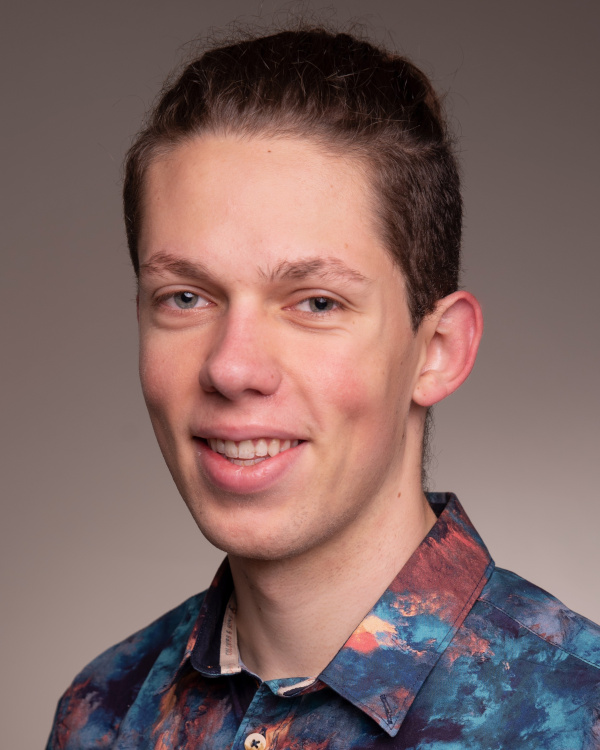}}]
{Tim Foißner} received the B.Sc. degrees in electrical engineering and information technology and in mechatronics from TU Darmstadt, Darmstadt, Germany, in 2023 and 2024, respectively, and the M.Sc. (Tech.) degree in automation and electrical engineering from Aalto University, Espoo, Finland, in 2026.

He is currently a Design Engineer with ABB Oy, Helsinki, Finland. His research interests include the design, analysis, modeling, and control of electric machines.
\end{IEEEbiography}

\begin{IEEEbiography}[{\includegraphics[width=1in,height=1.25in,clip,keepaspectratio]{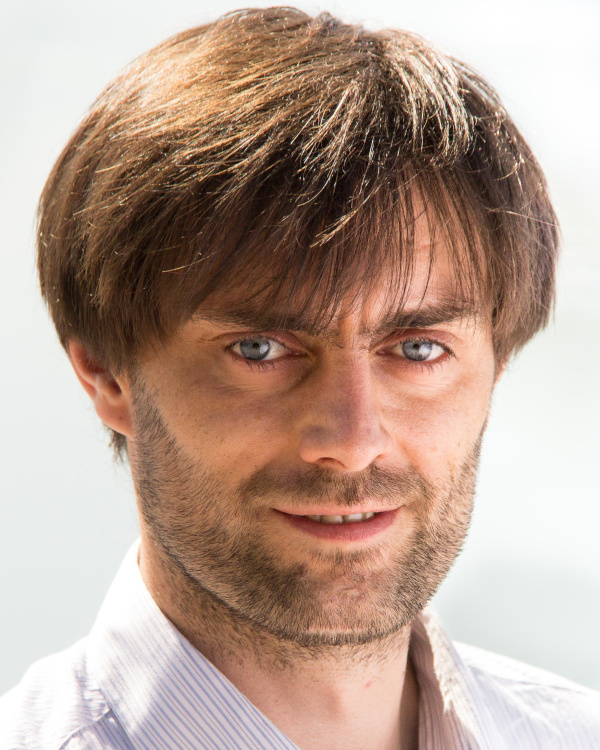}}]
{Floran Martin} received the Engineering Diploma in electrical engineering from Polytech Nantes, Nantes, France, in 2009, and the M.S. and Ph.D. degrees in electrical engineering from the University of Nantes, Nantes, in 2009 and 2013, respectively.

In 2014, he joined the Department of Electrical Engineering and Automation, Aalto University, Espoo, Finland, where he is currently a Staff Scientist. His research interests include modeling of magnetic materials as well as analyzing, designing, and controlling electric machines.
\end{IEEEbiography}

\begin{IEEEbiography}[{\includegraphics[width=1in,height=1.25in,clip,keepaspectratio]{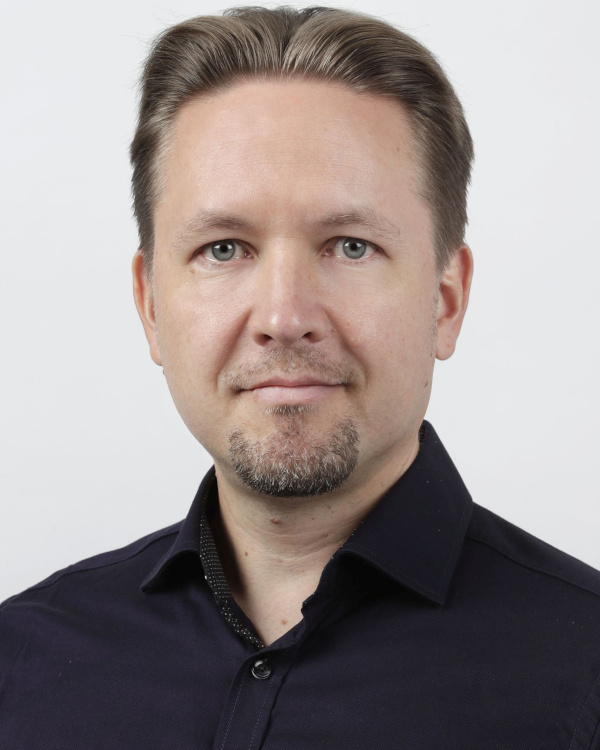}}]
{Antti Piippo} received the M.Sc. (Eng.) and D.Sc. (Tech.) degrees in electrical engineering from the Helsinki University of Technology, Espoo, Finland, in 2003 and 2008, respectively. 

He is currently a Corporate Executive Engineer with ABB Oy, Drives, Helsinki, Finland. His main research interests include the control of electric drives. 
\end{IEEEbiography}

\begin{IEEEbiography}[{\includegraphics[width=1in,height=1.25in,clip,keepaspectratio]{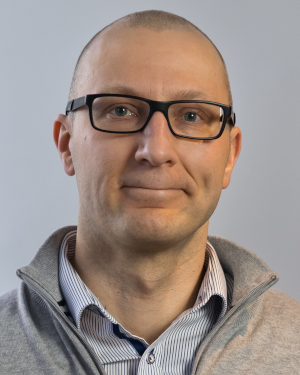}}]
{Marko Hinkkanen} (Fellow, IEEE) received the M.Sc. (Eng.) and D.Sc. (Tech.) degrees in electrical engineering from the Helsinki University of Technology, Espoo, Finland, in 2000 and 2004, respectively.
 
He is currently a Full Professor with the School of Electrical Engineering, 
Aalto University, Espoo, Finland. His research interests include control systems, physics-informed machine learning, electric machine drives, and power converters.
 
Dr. Hinkkanen was the recipient of eight paper awards, including the 2016 International Conference on Electrical Machines (ICEM) Brian J. Chalmers Best Paper Award, and the 2016 and 2018 IEEE Industry Applications Society Industrial Drives Committee Best Paper Awards. He was the corecipient of the 2020 SEMIKRON Innovation Award. He was the General Cochair of the 2018 IEEE 9th International Symposium on Sensorless Control for Electrical Drives (SLED). He is an Associate Editor of \textsc{IEEE Transactions on Power Electronics}.
\end{IEEEbiography}

\end{document}